\documentclass[aps,prb,reprint,superscriptaddress,nobibnotes,showkeys,tightenlines]{revtex4-1}
\usepackage{amsmath}
\usepackage{bbm}
\usepackage{commath}
\usepackage{graphicx}
\usepackage{color}
\usepackage{indentfirst}
\usepackage{setspace}
\usepackage{tabularx}
\usepackage{multirow}
\usepackage{subfigure}
\usepackage{float}
\usepackage{ragged2e}
\usepackage{natbib}
\usepackage{titlesec}
\titlespacing*{\section}{0pt}{0.5\baselineskip}{0.5\baselineskip}

\begin{document}

\title{A melting temperature database and \\a neural network model for melting temperature prediction} 

\author{Qi-Jun Hong}
\email[e-mail:]{ qhong@alumni.caltech.edu}
\affiliation{School for Engineering of Transport, Energy and Matter, Arizona State University, Tempe, AZ 85287, USA}

\date{\today}

\begin{abstract}
I build a melting temperature database that contains approximately 10,000 materials. Based on the database, I build a machine learning model that predicts melting temperature in seconds. The model features graph neural network and residual neural network architecture. The root-mean-square errors of melting temperature are 90 and 160K for training and testing, respectively. The model is deployed online and is publicly available.
\end{abstract}
\maketitle

High-performance refractory materials \cite{Padture2002,Wuchina2007,Perepezko2009,Lu2010,Liu2013} play an important role in applications ranging from gas turbines to heat shields for hypersonic vehicles. High melting temperature often correlates with desirable thermodynamic and mechanical properties, e.g., high-temperature materials strength, good ablation, creep resistance, etc. Unfortunately, the calculation and prediction of melting temperature is much more expensive and complex than many low-temperature properties, because the process involves the liquid phase and a huge amount of configuration sampling.

Previously, I built a fully automated tool, Solid and Liquid in Ultra Small Coexistence with Hovering Interfaces (SLUSCHI), for density function theory (DFT) melting temperature calculation \cite{Hong2013,Hong2016,Hongthesis}. I utilized this tool to predict the material, a Hf-C-N alloy, with the world’s highest melting temperature \cite{HfCN2015}, and the computational discovery was later partially confirmed from experiment \cite{Manara2016, Buinevich2020}. I have employed this tool to study hundreds of materials \cite{Hong2015, Ta_EOS2015, Kapush2017, Hong2019, Ushakov2019, Fyhrie2019, Hong2021}. The tool is freely distributed, and widely used in the community \cite{Addington2016,Guren2017}. It has been downloaded for more than 3,000 times. To the best of my knowledge, this computational method is one of the best approaches to calculate melting temperatures both accurately and relatively quickly via DFT. 

\twocolumngrid

Despite the progress, there remain challenges and hurdles to achieve automated materials discovery of high-melting-point materials. High-temperature properties are considerably more difficult and expensive to predict. At the pace of several days per material, the speed of DFT melting temperature calculation is very limited, which in turn put a serious constraint on its predictive power. We also rely heavily on human knowledge and intuition to design materials, e.g., the combinations and compositions of elements, which is constrained by our current understanding of materials. 

In this work, I aim to integrate DFT and machine learning (ML), which complement each other in terms of speed and accuracy, with an ultimate goal to build a system that is both robust and fast and that learns and improves by itself and truly achieves artificial intelligence. The ML model allows me to rapidly estimate melting temperatures, at a speed on the order of one second per material, while the DFT calculation provides robustness and accuracy at a much higher cost on the order of several days per material. 

\begin{figure*}[t]
\includegraphics[width=1.05\textwidth]{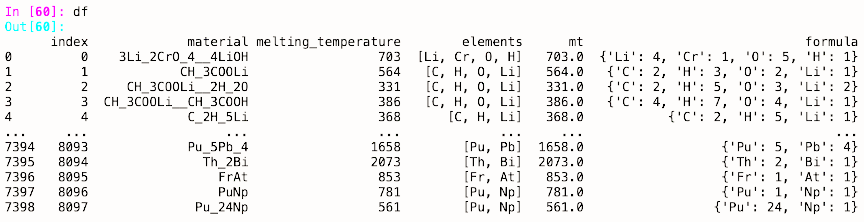}
\caption{\label{database}A screenshot of the melting temperature database I have built so far. The database currently contains 9375 materials, out of which 982 compounds are high-melting-temperature materials with melting points above 2000K.}
\end{figure*}

In order to utilize ML methods to predict melting temperature, I first build a melting temperature database via web crawling. 
Melting temperatures are collected and included in my database from the ASM Alloy Phase Diagram Database, as well as the ChemNet Database from the Department of Chemistry at the Moscow State University. DFT melting temperature calculations are included in the database as well.
My current melting temperature database contains 9375 materials, out of which 982 compounds are high-melting-temperature materials with melting points above 2000 Kelvin. 
The database consists of chemical composition, i.e., elements and concentration, or equivalently chemical formula, of the materials, and their corresponding melting temperatures, as shown in a screenshot of the database in Fig. \ref{database}.

\begin{figure*}[t]
\includegraphics[width=1.05\textwidth]{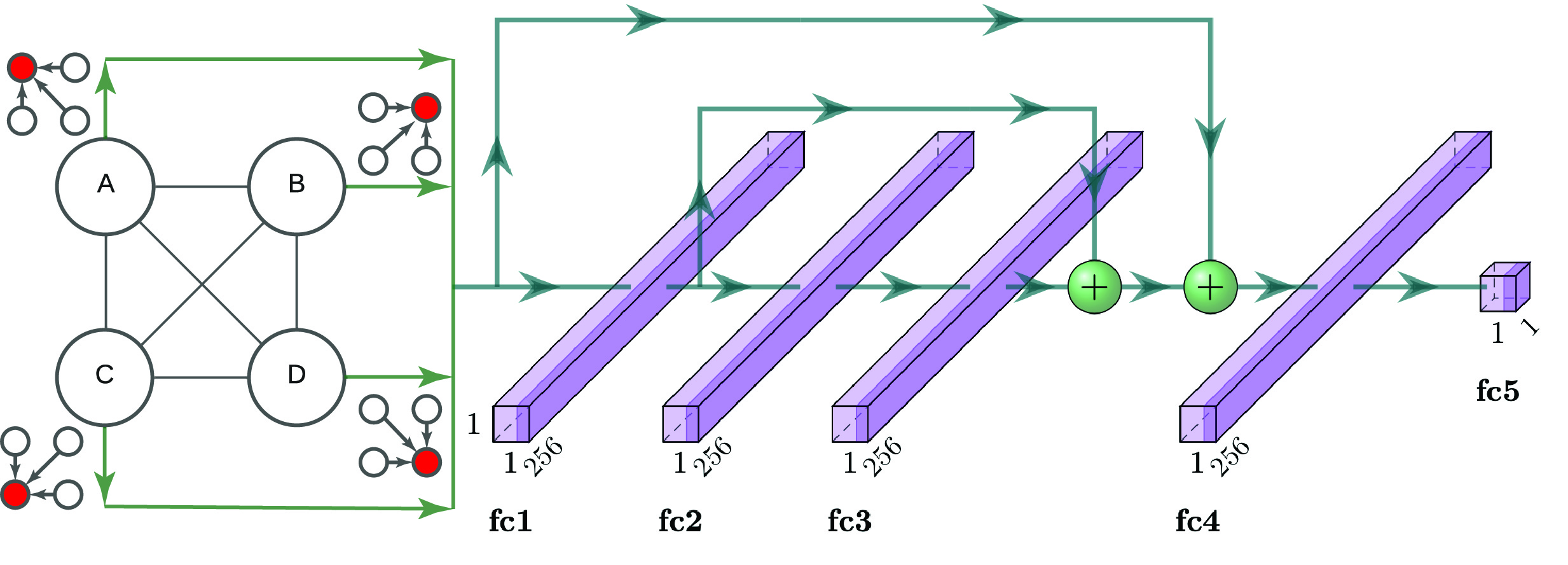}
\caption{\label{arch}Architecture of the GNN model for ML melting temperature prediction. Up to four elements and compositions are connected in this graph. Elements and compositions are first converted to features (such as atomic radius, electronegativity, valence electron, position in the
periodic table, etc.) and then connected in a GNN to exchange information. The output of the GNN encoders is then fed into a ResNet for melting temperature regression.
}
\end{figure*}

Based on the database, I then build a machine learning neural network model to predict melting temperature.  I employ the Graph Neural Networks (GNN) \cite{Scarselli2009} architecture, as well as residual neural network (ResNet) \cite{He2016}, as illustrated in the architecture in Fig. \ref{arch}. 
When a material, i.e., its elements and compositions, is fed to the neural network, 
each element, that constitutes the material, is first converted to 14 features, such as atomic radius, electronic negativity, electron affinity, valence electrons, position in the periodic table, etc. These features are encoded and passed to the next layer, a process I expect as the individual contribution to the melting temperature from each element.
In addition, elements communicate to each other via the GNN architecture, thus leading to impacts from the binary and ternary combination of the elements. These encoded impacts are passed to the next layer as well.
This layer, consisting of singular, binary, and ternary interaction of the elements and compositions of the material of interest, is fed into a 4-layer ResNet, which leads to the regression of the melting temperature.
Currently the number of elements are limited to four, but this constraint can be removed if more combination of elements are needed.
The GNN architecture undergoes two iterations of communication among elements, as I find further rounds do not improve performance of the model.

\begin{figure}
\includegraphics[width=0.49\textwidth]{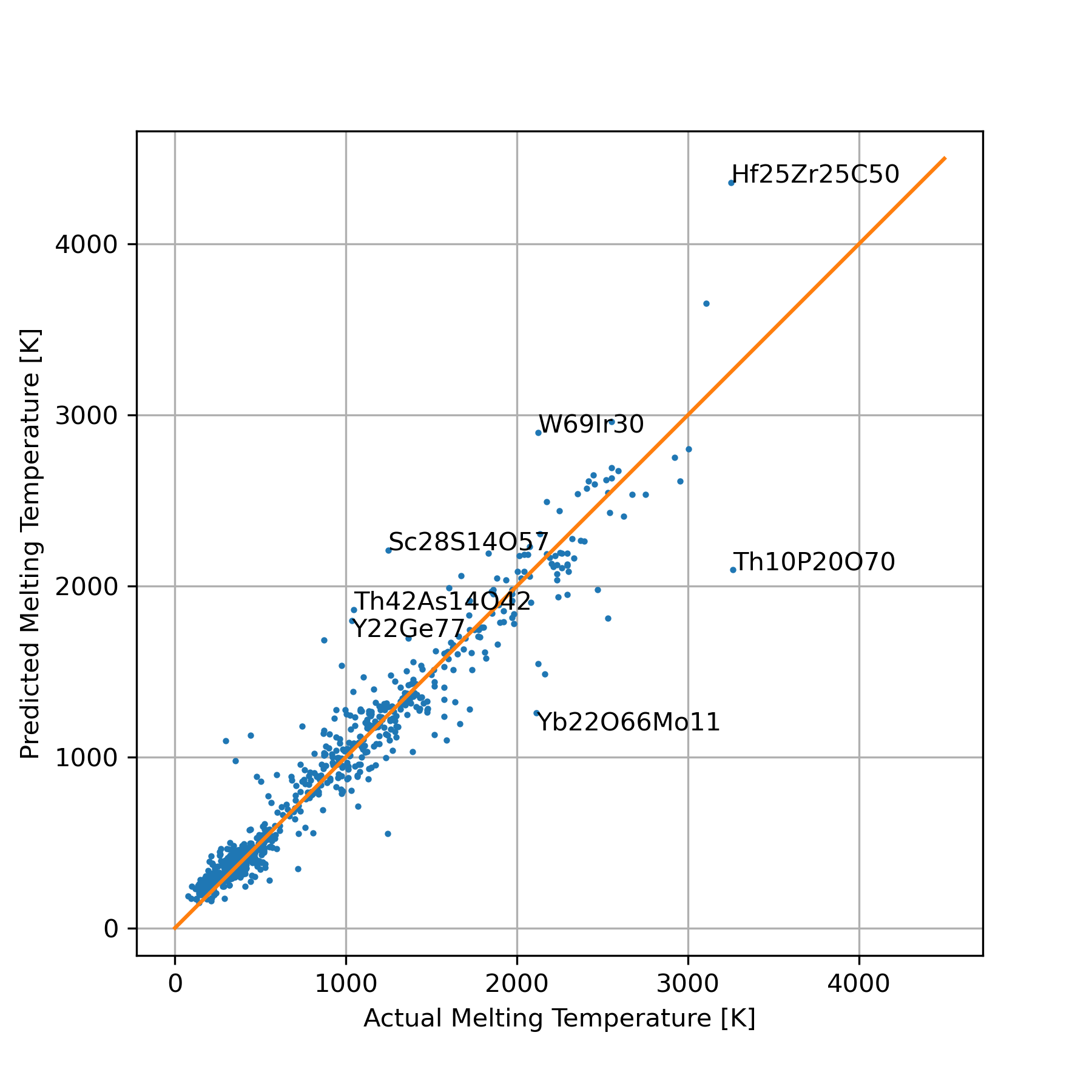}
\caption{\label{results} Predicted vs. actual melting temperatures in the testing dataset. The root-mean-square errors of melting temperature are 90 and 160 K for the training and testing sets, respectively. Compounds with large errors are labeled for further investigation.
}
\end{figure}

The 9375 materials are randomly assigned to training and testing sets, with 8635 materials in the training set, and 740 materials for testing. 
The neural network model is built with Tensorflow \cite{tensorflow}. The training process takes approximately 1000 epochs of optimization.
The root-mean-square errors of melting temperature are 90 and 160 K for the training and testing sets, respectively, as shown in Fig. \ref{results}.
At first glance, the errors are impressively small. After all, this is a machine learning model based on neural networks, but the accuracy appears to be comparable to much more complex and expensive methods, such as density functional theory calculations. However, I note that while this model performs well in known region, it may lose robustness in unexplored regions, making it inferior to DFT calculations. I will discuss later future plans to incorporate the ML model and DFT to complement the two methods. 

I deploy the model on my research webpage \cite{ML_model_webpage}, so other researchers in the community may use the Application Programming Interface (API) to estimate the melting temperatures for the materials of their interests. 
The model is currently hosted on Microsoft Azure, and it will be hosted at the ASU Research Computing center in the future. 
To use the model, the user needs to visit the website and input the chemical compositions of the materials of interest. The model will respond with a predicted melting temperature in seconds, as well as the actual melting temperatures of the ``nearest neighbors" in the database.

In order to further improve this model, I will need to collect considerably more data and significantly increase the size of the database. 
My DFT computational tool, SLUSCHI, complements this ML model on this purpose, as it generates accurate DFT melting temperatures automatically on regular basis.
As more data being included in the melting point database, it will further improve the accuracy of the ML melting point predictor.

An integration of the ML model and the DFT tool will allow me to build an iterative and self-consistent framework for materials design and discovery of high melting temperature materials.
The fully automated computational method and tool allow me to generate data for my melting point database, in addition to the experimental data I collect. This potentially unlimited source of DFT data enables me to systematically improve my ML model.
Furthermore, the ML model rapidly screens potential promising materials and sends them to the DFT tool for verification.
This mutual communication between the ML and DFT tools makes this ML-DFT framework unique and promising to achieve the goal of using artificial intelligence for materials design and discovery of high-melting-temperature materials.

To summarize, I build a melting temperature database and a machine learning model for melting temperature prediction. 
The database so far contains approximately 10,000 materials, and the ML model achieves high accuracy with 90 and 160K RMSE  of melting temperature for training and testing, respectively.
The model is deployed online and publicly available.
Future integration of the database, the ML model, and the DFT tool will create a framework for materials design and discovery based on melting temperature related properties.

\section*{Acknowledgements}
The author thanks Drs. Sergey V. Ushakov, Axel van de Walle, Alexandra Navrotsky, and Yang Jiao for discussions.

\bibliography{citepapers}

\end{document}